\documentstyle[amssymb,12pt,epsfig]{article}

\input{tcilatex}

\begin{document}

\noindent{\large Fusion and breakup of halo nuclei\bigskip }

\noindent M.S. Hussein$^{a}$, L.F. Canto$^{b}$ and R. Donangelo$^{b}\bigskip 
$

\noindent $^{a}$Instituto de F\'{i}sica, Universidade de S\~{a}o Paulo,

\noindent C.P. 66318, 05389-970, S\~{a}o Paulo, SP, Brazil\bigskip 

\noindent $^{b}$Instituto de F\'{i}sica, Universidade Federal do Rio de
Janeiro,

\noindent C.P. 68528, 21945-970, Rio de Janeiro, RJ, Brazil\vspace{1cm}

We discuss the effect of the coupling to the break-up channel on the total
fusion involving a halo nucleus with a heavy target. We show that there is a
competition between the hindrance arising from this coupling mostly at above
barrier energies, and the enhancement which ensues\ at sub-barrier energies
owing to the static effect of the extended matter distribution and the
coupling to the soft modes.\bigskip

\noindent {\bf 1. INTRODUCTION\smallskip }

The fusion of heavy ions is of paramount\ importance in astrophysics and in
the production of super heavy elements\ (SHE). With the advent of secondary
beams of neutron and proton rich nuclei, it is important to assess\ how the
complete fusion of these nuclear species behave \ as a function of
bombarding energy especially near the Coulomb barrier [1]. Several of the
light neutron and proton rich nuclei exhibit halo structures, with a compact
core plus one or two loosely\ bound nucleons occupying a far away orbit.
Systems such is $^{11}Li$ and $^{6}He$ are two-neutron, borromean halo
nuclei, while $^{11}Be$ and $^{19}C$ are one-neutron halo nuclei. The
isotope $^{8}B$ has been confirmed to be a one-proton halo while $^{17}F$ is
a normal nucleus in its ground state but acquires a one-proton halo nature
in its 1st excited state. We ask the question of how the above systems fuse.
One important feature to remember about these loosely bound systems is their
collective response. They exhibit the so-called soft giant resonances (pygmy
resonances), the most notorious of which is the soft dipole resonance, very
nicely confirmed in $^{6}$He by Nakayama et al [2]. On the other hand the
threshold for break-up is very small $(<\,1$ $MeV)$, making the study of the
fusion of these nuclei with heavy targets an interesting endeavour, since
the coupling to the soft modes tend to enhance fusion while the coupling to
break-up reduces fusion. The latter is true since, if the Coulomb and
nuclear coupling is of long range and strong, which is the case when fusion
occurs with a heavy nucleus, then the projectile (the halo nucleus) would
break before it reaches the target. This led us to propose a model for
complete fusion which looks like

\begin{equation}
\sigma _{CF}=\frac{\pi }{k^{2}}\sum_{i}\left( 2\ell +1\right) T_{\ell
,i}\left( E\right) P_{\ell ,i}\left( E\right)
\end{equation}

\newpage
\topmargin 6 pt
\textheight 230mm
\textwidth 160mm

\noindent where $T_{\ell ,i}\left( E\right) $ is the fusion transmission coefficient
for the $l-th$ partial wave and $P_{\ell ,i}\left( E\right) $ is the
break-up survival probability [3,4,5]. The sum in Eq.(1) is over all bound
states in the projectile and target. Simple approximations were then used
for $T_{\ell ,i}\left( E\right) $\ and $P_{\ell ,i}\left( E\right) $ and a
few examples were considered. The overall result from this model was a
reduction of $\sigma _{CF}$ near and above the barrier energy and an
enhancement below. This trend was later confirmed by Takigawa et al [6] and
more recently by the coupled channel calculation with continuum
discretization of Hagino et al [7].

Though our\ results were\ based on intuitive\ arguments, one can derive\ Eq.
(1) from general reaction theory. The same theory [8] allows the obtention
of the incomplete fusion cross-section $\sigma _{ICF}$, which in many
instances is not easy to distinguish from $\sigma _{CF}$. As an example we
take the fusion of $^{6}$He recently measured by Trotta et al [9] on $%
^{238}U $ and Kolata et al [10] on $^{209}Bi$. Here the incomplete fusion is
that where $^{6}He$ is broken up and $^{5}He$ or $^{4}$He is absorbed by the
target. The remaining neutrons when detected are difficult to distinguish
from evaporation neutrons from the compound nucleus.

In the present contributions we give a brief review of theoretical attempts
to calculate the fusion of halo nuclei with heavy targets. We also present a
short review of the experimental data currently available.\bigskip

\noindent {\bf 2. EXPERIMENTAL DATA\smallskip }

So far, four measurements were made on the fusion of $^{11}Be$ and $^{6}He$
with heavy targets at near-barrier energies. Signorini et al [11] reported
the measurements of the fusion systems $^{9,10,11}Be+^{209}Bi$ at energies
close to the Coulomb barrier. These authors reach the conclusion that in the
weakly bound normal nucleus $^{9}$Be, the fusion with $^{209}Bi$ is found
significantly reduced at above barrier energies, owing to the coupling to
the breakup channel. This supports our discussion in the introduction
concerning the irreversible nature of the coupling to the break-up
continuum. The cases of $^{10}Be$ and $^{11}Be$ are more subtle to
understand. However, the more recent results of [9] on $^{6}He+^{238}U$ and
[10] on $^{6}He+^{209}Bi$ do indicate that at below the barrier energies the
fusion cross section is enhanced. Thus the halo shows itself as enhancement
at sub-barrier energies and the break-up hinders\ the fusion; the effect
becoming more important at above barrier energies. We should of course
remind the readers that $^{6}He$ is a borromean 2n-halo nucleus while $%
^{11}Be$ is one-neutron halo nucleus. Signorini at al [11] found that the,
normal, strongly bound $^{10}Be$ isotope presents a larger fusion cross
section than that of $^{11}Be$ at sub-barrier energies. This effect does not
fit into the picture we have just presented. We should mention that is the
case of the comparison between $^{6}He$ and $^{4}He$ fusion with $^{209}Bi$
by Kolata et al [10], the effects of the halo (enhancement) at sub-barrier
energies, were quite evident.

The fusion of the proton rich isotope $^{17}F$ with $^{208}Pb$ was measured
by Rehm et al [12]. This weakly bound nucleus has a normal ground state, but
its first excited state is mostly $\ell =0$ and seems to exhibit halo
features. The results of Ref. (12) indicate a rather normal behaviour of the
complete fusion cross-section, with a very small effect due to break-up,
though the break-up channel coupling $\left( ^{17}F\rightleftarrows
^{16}O+p\right) $would seem to reduce the Coulomb barrier, contrary to the
case of the fusion of neutron-rich, drip-line, nuclei.

If there is strong break-up effect on the fusion cross-section that leads to
its reduction then the break-up\ cross-section itself should be noticeable.
In the recent measurement by Hinde \ et al [13], of $^{9}$Be+$^{208}$Pb at
near-barrier energies the elastic break-up cross-section was measured. Here
the prompt break-up\ channel is $\alpha +\alpha +n$ with a $Q=1.57\,MeV$.
The authors reach the conclusion that this channel is responsible for the
large suppression of the complete fusion cross-section at above-barrier
energies (one-dimensional barrier calculation has to be multiplied by 0.68
to account for the observation).

This reduction in the complete fusion seems to be accompanied by a large
incomplete fusion cross-section where one of the $\alpha $-particles fuses
with $^{208}$Pb. Hinde et al [15] found that the incomplete fusion
probability is about $0.32\pm 0.07$, clearly attesting for the unitarity
constraint that $P_{CF}+P_{ICF}$ should be one. The corresponding total
fusion cross-section $\sigma _{F}\equiv \sigma _{CF}+\sigma _{ICF}$ should
be accountable by the simple one-dimensional barrier penetration model.
These findings corroborates the earlier results of Dasgupta et al [14] which
showed a considerable hindrance in $\sigma _{F}$ at above barrier energies.
Hinde et al [13] also discuss the dependence of the fusion hindrance factor
on the charge of the target. They found that $P_{CF}\,$depends strongly on
the target charge. In the experiment of Ref. [15] involving $^{6,7}Li$
fusion with $^{9}Be$ and $^{12}C$ targets, it was reported that suppression
of up to 70\% is found. This, however, was contested recently [16,17]
through independent measurements of several of the systems studied in Ref.
[15]. The conclusions of [16,17] is that there is no supression of $\sigma
_{CF}$ for these light systems.

In fact, the 70\% fusion hindrance in the fusion of $^{6}Li$ and $^{7}Li$,
was indeed found in Ref. [18] but on a heavy target, $^{209}Bi$. We can use
Eq. (1) as a guide and take for the break-up survival probability the simple
DPP form [4] with the Coulomb DPP proportional to $Z_{T}^{2}$ [19,20]. If
the effect of the survival probability on the fusion of $^{7}Li+^{209}Bi$ is
75\% [18] at near-barrier energies, then, for the $^{7}Li+^{12}C$ case of
[15]. the effect is tremendously reduced since the quantity $\left(
Z_{C}/Z_{Bi}\right) ^{2}\simeq 0.005$, implies an over all reduction in $%
\sigma _{CF}$ for $^{7}Li+^{12}C$ of about $\exp \left( -0.005\,\ell n\left(
1/0.75\right) \right) =0.99989$, namely no reduction at all. Of course there
are the nuclear break-up effects, but these are very small as well.
Accordingly there is no suppression of $\sigma _{CF}$ for light systems, in
total agreement with the conclusion of Refs. [16,17].

Of course other processes such as fast\ fission, also results in a
considerable hindrance in the complete fusion involving heavy targets [21].
This is similar to the effects of Deep Inelastic Collision on $\sigma _{CF}$
in the so-called Region II [1].\bigskip

\noindent {\bf 3. THEORETICAL MODELS}\smallskip

Theoretical description of tunnelling phenomena, such as sub-barrier fusion,
in the case of strong channel coupling is made with recourse to coupled
channels theory [22]. This theory has been extensively and successfully used
to describe the sub-barrier fusion of stable nuclei, which invariably
exhibits enhancement over the simple one-dimensional barrier penetration
model [1]. The channels that are taken into account correspond to bound
states of the partners. In extending this picture to the fusion of halo
nuclei, one must take into account the break-up channel coupling. This has
been done is the past for deuteron scattering from different targets within
the so-called Continuum Discretized Coupled Channels (CDCC) Method [23]. The
effect of the deuteron break-up on the elastic scattering was found to be
representable by a weaker attraction, and absorption (implying that the
dynamic polarization potential, DPP, associated with the deuteron break-up
has a repulsive real part and attractive imaginary part (absorption)
[4,5,23,24]). It was from these findings that we were lead to use Eq. (1)
[4,5,24]\ which later modified to incorporate the real part of the DPP
[3,24] in the survival probability, $P_{\ell ,i}\left( E\right) $. In a
later publication, Dasso and Vitturi [25] used an effective one bound
``break-up'' channel and reached the conclusion that the break-up leads to
enhancement of fusion. In a later paper, these authors [7] modified their
work by adding, many bound ``break-up'' channels and reached a similar
conclusion as ours, namely, break-up hinders fusion at above-barrier
energies and enhances it at lower energies. A more quantitative comparison
shows that the shape of their cross section is somewhat different from ours
at the barrier region and the transition from enhancement to hindrance
occurs at a slightly higher energy.

In fact, the effect of the widths of these resonances (excited states with
finite life times) was studied in Refs. [26,27]. The overall effect of the
width as compared to bound excited state, is a reduction in fusion. However,
at very low energies, the pygmy resonances act as if they were bound excited
states and thus the enhancement seen in the sub-barrier fusion calculation
of Ref. [7] must arise from the above effects, as well as the farther
extension of the matter distribution, and not to break-up.

Quite\ recently, A. Diaz-Torres and I.J. Thompson [28] performed the most
complete calculation of the fusion cross-section of halo nuclei using the
CDCC method. They\ found a large reduction in the complete fusion both above
and below the barrier for the system $^{11}Be+^{208}Pb$. They also
calculated the incomplete fusion cross-section as that corresponding to flux
loss from the break-up channels (the continuum is discretized into several
channels). This definition of $\sigma _{ICF}$ which follows that of Hagino
et al [7] is an overestimate since the incomplete fusion is the fusion of
the heavy charged fragment only. See Ref. [8], for more details. In fact in
many of the papers found in the literature on fusion of two-cluster (or
three-cluster) nuclei, there is always ambiguities in actually measuring and
also calculating $\sigma _{ICF}$. An exception to this is found in the work
of Hinde et al [13] where $\sigma _{ICF}$ \ was correctly measured for the
system $^{11}Be+^{209}Bi$.

It is of interest to analyze the general structure of the total reaction
cross section, $\sigma _{R}$, of the system studied by A. Diaz-Torres and
Thompson [28], namely $^{11}Be+^{208}Pb$ at low energies. From the formal
analysis of Hussein [29], we can write the following expression for the
reaction cross-sector

\begin{equation}
\sigma _{R}=\sigma _{DIR}+\sigma _{F}
\end{equation}
where $\sigma _{DIR}$ is the total direct cross section which contains, in
the specific case of $^{11}Be+^{208}Pb$ at very low energies, just the
inelastic, $\sigma _{INE}$ (mostly Coulomb) excitation of the $1/2^{-}$\
state at 0.32\ MeV and the elastic break-up cross-section corresponding to $%
^{11}Be$ splitting into $^{10}Be$ plus a neutron in addition to the
situation where the neutron is absorbed by the target forming $^{209}Pb$,
namely the one-neutron removal cross-section $\sigma _{-n}$. The component $%
\sigma _{F}$ is the total fusion cross-section alluded to earlier, which
contains the complete fusion $\left( ^{11}Be+^{208}Pb\rightarrow
^{\;219}Rn\right) $ plus the incomplete fusion $\sigma _{ICF}$ where $%
^{10}Be $ is captured by the target It is more likely that $\sigma _{ICF}$
should be just $^{10}Be+^{208}Pb\rightarrow ^{\;218}Rn\;$since $^{10}Be$ is
long-lived in the present context. Thus $\sigma _{ICF}=\sigma _{-\;^{10}Be}$.

Accordingly we can write

\begin{equation}
\sigma _{DIR}=\sigma _{INE}+\sigma _{bup}+\sigma _{-n}
\end{equation}
and 
\begin{equation}
\sigma _{F}=\sigma _{CF}+\sigma _{^{-\,\;10}Be}
\end{equation}

In the calculation of Hagino et al [7] and of Diaz-Torres and Thompson [28],
the total fusion cross-section, $\sigma _{F}$ was obtained from 
\begin{equation}
``\sigma _{F}"=\sigma _{CF}+\sigma _{^{-\,\,10}Be}+\sigma _{-n}\text{ \ ,}
\end{equation}
or 
\begin{equation}
``\sigma _{F}"\equiv \sigma _{CF}+``\sigma _{ICF}"
\end{equation}

Therefore $``\sigma _{ICF}"$ is an overestimate of the incomplete fusion
cross-section. Further, if $\sigma _{DIR}$ is ignored, as done in both
references, then unitarity would require an underestimate of $\sigma _{CF}$.
This explains the findings of [28] and points to ways of correcting them.
One should resort to three - or four - body calculations in order to
calculate $\sigma _{-n}$ and $\sigma _{^{-\;10}Be}$. Short of a full
fledged\ Faddeev (3-bodies) or Jakubovskij\ (4-bodies) treatments of the
reactions of one - and two - nucleon halo nuclei, respectively, one resorts
to approximate treatments. In Ref. [8], using the Hussein-McVoy [30]
formalism for inclusive break-up reactions and improvements on it,\ it was
find that one may calculate the removal cross-section of the heavy fragment, 
$\sigma _{^{-\;10}Be}$, according to the practical formula [8] 
\begin{equation}
\sigma _{ICF}=\sigma _{^{-\;10}Be}=\frac{\pi }{k_{^{10}Be}^{2}}\frac{%
v^{\prime }}{v}\sum\limits_{\ell _{1}}\left( 2\ell _{1}+1\right) T_{\ell
_{1}}^{(^{10}Be)}\left( E_{1}\right) \left[ 1-P_{\ell _{o}}\left( E\right) %
\right] \left[ 1-T_{\ell _{2}}^{(n)}\left( E_{2}\right) \right]
\end{equation}

\noindent where $\left[ 1-P_{\ell _{o}}\left( E\right) \right] $ is the
break-up probability of the projectile, $T_{\ell _{1}}^{(^{10}Be)}\left(
E_{1}\right) $ is the fusion probability for $^{10}Be$ and $\left[ 1-T_{\ell
_{2}}^{(n)}\left( E_{2}\right) \right] $ is the survival probability of the
neutron in the process $^{11}Be\rightarrow ^{10}Be+n$. In Eq. (7) $v$ and $%
v^{\prime }$ are the relative velocities in the elastic and break-up
channels. The calculation of $\sigma _{^{-\,10}Be}$ is presently in
progress. The above considerations should be worked out within a few-body
treatment of fusion [31]. However, $\sigma _{ICF}$, according to Eq. (7),
may be calculated within the CDCC of Ref. [28].

One last comment concerning the complete fusion cross section calculated
within CDCC by Diaz-Torres and I.J. Thompson [28] for $^{11}Be+^{208}Pb$.
There is a reason to expect another mechanism that would further enhance the
very low values of $\sigma _{CF}$. After discretizing the continuum, and
thus replacing the density of states of the two-clusters continuum by a much
smaller density of discrete states, one looses part of the fusion flux. This
lost flux, should be accounted for within CDCC by an appropriate incoherent
addition of a fluctuation fusion contribution. Ref. [32] discusses in fact
the statistical nature of the coupling to the continuum, which would
require, besides the ``average'' CDCC , the statistical treatment of the
fluctuation, just as is done in the description of DIC [33]. This should
increase the Diaz-Torres and Thompson [28] complete fusion cross-section.
Work along this line is in progress.\bigskip

\noindent {\bf 4. CONCLUSIONS\smallskip }

We have presented in this short review an account of the present status of
the activities in the field of the fusion of halo nuclei. We have discuss
the results of the available data and critically assessed some of the
interpretations. We have also discuss theoretical attempts made so far to
calculate the complete and incomplete fusion cross-sections. In particular
we went at length in considering the effect of the coupling to the break-up
channel. The overall conclusion that seems to emerge from several of these
studies is that this coupling always leads to a reduction in $\sigma _{CF}$
above the barrier. At sub-barrier \ energies, other features of halo nuclei
(extended mass distribution, pygmy resonances) come into play and they lead
to an enhancement of $\sigma _{CF}$ compared to the single channel model
results. We have also analysed the recent calculation of Diaz-Torres and
Thompson [28] within the CDCC method and pointed to ways of improving the
theory.

Fusion and break-up of halo nuclei continues to be a challenging field and
certainly deserves further experimental and theoretical studies.\bigskip

\noindent {\bf ACKNOWLEDGMENT\medskip }

This work was supported in part by FAPESP and the CNPq.\bigskip

\noindent {\bf REFERENCES\medskip }

\begin{enumerate}
\item  For a recent review see, C.A. Bertulani, M.S. Hussein and G.
M\"{u}nzenberg, ``Physics of Radioactive Beams'' (Nova Science, New York,
2001), Chapter 12.

\item  S. Nakayama et al. Phys. Rev. Lett 85 (2002) 262.

\item  M.S. Hussein, M.P. Pato, L.F. Canto and R. Donangelo, Phys. Rev. C46
(1992) 377; Phys. Rev. C47 (1993) 2398.

\item  L.F. Canto, R. Donangelo, P. Lotti and M.S. Hussein, Phys. Rev. C52
(1995) R2848; L.F. Canto et al. J. Phys. G23 (1997) 1465; W.H.Z.
C\'{a}rdenas et al., Nucl. Phys. A703 (2002) 673.

\item  M.S. Hussein et al., Nucl. Phys. A588 (1995) 85C.

\item  N. Takigawa, M Kuratani and H. Sagawa, Phys. Rev. C47 (1993) R2470.

\item  K. Hagino, A. Vitturi, C.H. Dasso and S. Lenzi, Phys. Rev. C51 (2000)
037602.

\item  L.F. Canto et al., Phys. Rev. C58 (1998) 1107.

\item  M. Trotta et al., Phys. Rev. Lett. 84 (2000) 2342.

\item  J.J. Kolata et al., Phys. Rev. Lett. 81 (1998) 4580.

\item  C. Signorini et al., Eur. Phys. J. A5 (1999) 7; C. Signorini, Eur.
Phys. J. A13 (2002) 129.

\item  K.E. Rehm et al., Phys. Rev. Lett. 81 (1998) 3341.

\item  D.J. Hinde et al., Phys. Rev. Lett. 89 (2002) 272701-1.

\item  M. Dasgupta et al., Phys. Rev. Lett. 82 (1999) 1395.

\item  J. Takahashi et al., Phys. Rev. Lett. 78 (1997) 30; A. Szanto de
Toledo et al., Nucl. Phys. A679 (2000) 175.

\item  A. Mukherjee et al., Nucl. Phys. A635 (1998) 205; A. Mukherjee et
al., Nucl. Phys. A645 (1999) 13; A. Mukherjee et al., Phys. Ltett. B526
(2002) 295.

\item  R.M. Anjos et al., Phys. Lett. B534 (2002) 45.

\item  M. Dasgupta et al., Phys. Rev. C66 (2002) 041602 (R).

\item  M.V. Andr\'{e}s, J. G\'{o}mez-Camacho and M.A. Nagarajan, Nucl. Phys.
A579 (1994) 273.

\item  L.F. Canto, R. Donangelo, P. Lotti and M.S. Hussein, Nucl. Phys. A589
(1995) 117.

\item  D.J. Hinde, M. Dasgupta and A. Mukherjee, Phys. Rev. Lett. 89 (2002)
282701.

\item  See, e.g. M. Dasgupta et al., Ann. Rev. Nucl. Part. Sci. 48 (1998)
401; G.R. Satchler et al., \ Ann. Phys. 178 (1987) 110; A.M.S. Breitschaft
et al., Ann. Phys. 243 (1995) 420.

\item  N. Anstern et al., Phys. Rep. 154 (1987) 110.

\item  L.F. Canto et al., Nucl. Phys. A542 (1992) 131.

\item  C.H. Dasso and A. Vitturi, Phys. Rev. C50 (1994) R12.

\item  M.S. Hussein and A.F.R. de Toledo Piza, Phys. Rev. Lett. 72 (1994)
2693; M.S. Hussein et al., Phys. Rev. C51 (1995) 846.

\item  K. Hagino and N. Takigawa, Phys. Rev. C58 (1998) 2872.

\item  A. Diaz-Torres and I.J. Thompson, Phys. Rev. C65 (2002) 024606.

\item  M.S. Hussein, Phys. Rev. C30 (1984) 962.

\item  M.S. Hussein and K. W. McVoy, Nucl. Phys. A445 (1985) 124.

\item  B.V. Carlson, T. Frederico and M.S. Hussein, to be published.

\item  S. Drozdz, J. Okolowicz, M. Polszajazak and I. Rooter, Phys. Rev. C62
(2000) 024313.

\item  See, e.g. D. Agassi, C.M. Ko and H.A. Weidenm\"{u}ller, Ann. Phys.
(N.Y.) 107 (1977) 140; H. Feshbach, A. K. Kerman and S.E. Koonin, Ann. Phys.
(N.Y.) 125 (1980) 429.
\end{enumerate}

\end{document}